# $W$ Anomalous Moments and the Polarization Asymmetry Zero in $\gamma e \to W\nu$*

**Stanley J. Brodsky and Thomas G. Rizzo**

Stanford Linear Accelerator Center

Stanford University, Stanford, California 94309

and

**Ivan Schmidt**

Universidad Federico Santa María

Casilla 110–V, Valparaíso, Chile

ABSTRACT

We show from general principles that there must be a center of mass energy, $\sqrt{s_0}$, where the polarization asymmetry $A = \Delta\sigma(\gamma e \to W\nu)/\sigma(\gamma e \to W\nu)$ for circularly-polarized photon and electron beams vanishes. In the case of the Standard Model, the crossing point where the asymmetry changes sign occurs in Born approximation at $\sqrt{s_{\gamma e}} = 3.1583\ldots M_W \simeq 254$ GeV. We demonstrate the sensitivity of the position of the polarization asymmetry zero to modifications of the SM trilinear $\gamma WW$ coupling. Given reasonable assumptions for the luminosity and energy range for the Next Linear Collider(NLC) with a backscattered laser beam, we show that the zero point, $\sqrt{s_0}$, of the polarization asymmetry may be determined with sufficient precision to constrain the anomalous couplings of the $W$ to better than the 1% level at 95% CL. In addition to the fact that only a limited range of energy is required, the polarization asymmetry measurements have the important advantage that many of the systematic errors cancel in taking cross section ratios. The position of the zero thus provides an additional weapon in the arsenal used to probe anomalous trilinear gauge couplings.

*Work supported in part by Department of Energy contracts DE–AC03–76SF00515 and DE–AC02–76ER03069, and by Fondo Nacional de Investigación Científica y Tecnológica, Chile, contract 1931120.

## I. INTRODUCTION

Precision measurements of $Z$-pole observables at LEP and SLC [1] combined with the new $W$ mass determinations [2] and the discovery of the top quark at the Tevatron [3, 4] have demonstrated that the Standard Model (SM) provides an excellent description of physics below the electroweak scale. There are many reasons to believe, however, that new physics beyond the SM must exist, but it remains unclear just how or where it will first be directly observed. The scale of such new physics may not be far away, perhaps $\simeq 1$ TeV, in which case it will surely manifest itself at existing or planned colliders.

One of the most sensitive measures of new physics beyond the SM are the values of the electroweak moments of the various leptons, quarks, and gauge bosons. In the SM, the anomalous magnetic moments $\Delta\mu = (g-2)eS/2M$ of spin $S = \frac{1}{2}$ and $S = 1$ fundamental fields (with mass $M$)and the anomalous electric quadrupole moments $\Delta Q = Q + e/M^2$ of the vector bosons vanish at the tree level due to the requirements of gauge invariance and renormalizability, thus ensuring a quantum field theory which has maximally convergent high energy behavior. Deviations from these canonical values of the magnetic dipole and electric quadrupole moments beyond the usual SM radiative corrections [5] may reflect new physics or new interactions at high energies such as supersymmetry [5], technicolor, or compositeness [6].

Precision measurements at the $Z$-pole and elsewhere have already placed rather stringent restrictions on anomalous $(V = \gamma, Z)f\bar{f}$ couplings [7]. However, direct experimental probes of the trilinear gauge boson couplings, $VW^+W^-$, are still at a rather early stage [8, 9, 10]. If the energy scale of the new physics is indeed of order 1 TeV, it is anticipated on rather general grounds that these anomalous trilinear couplings can be no larger than $\mathcal{O}(10^{-2})$ [11]. Experiments have yet to achieve sensitivity at this level. However, it is ex-pected that the vector boson couplings will eventually be probed at the precision of 1% or better at high energy hadron and $e^+e^-$ colliders through processes such as $W$-pair production from fermion pair annihilation, $e^+e^- \to W^+W^-$, $q\bar{q} \to W^+W^-$, and associated production, $q\bar{q}' \to WZ^0, \gamma$.

The advent of backscattered laser beams at $e^+e^-$ colliders will allow tests of the anomalous couplings of the gauge bosons through measurements of the high energy photon collision processes $\gamma\gamma \to W^+W^-$, $\gamma\gamma \to Z^0Z^0$, $\gamma e \to Z^0\gamma$, and $\gamma e \to W\nu$. A distinctive feature of the $\gamma e \to W\nu$ process is that it isolates the on-shell photon $\gamma WW$ vertex in a model-independent manner. In this paper we shall show that measurements of the $\gamma e \to W\nu$ cross section with polarized photon and polarized electron beams can be used to test novel features of the canonical couplings of the $W$ and provide high precision measurements of its magnetic and quadrupole moments at a precision below 1%.

A remarkable consequence of the canonical couplings of fermions and gauge bosons in the SM is that the integral that appears in the Drell-Hearn Gerasimov sum rule(DHG) [12, 13] vanishes. This interesting observation was first made for quantum electrodynamics and also for the more general case of the SM by Altarelli, Cabibbo and Maiani [14]. Even more generally, one can use a quantum loop expansion to show [15] that the logarithmic integral of the spin-dependent part of the photoabsorption cross section, *i.e.*,

$$\int_{\nu_{th}}^{\infty} \frac{d\nu}{\nu} \Delta\sigma_{\text{Born}}(\nu) = 0 \tag{1}$$

for any $2 \to 2$ SM process $\gamma a \to bc$ at the Born level. The particles $a, b$ and $c$ are *arbitrary* (so long as $a$ carries non-zero spin!) and can be identified as leptons, photons, gluons, quarks, elementary Higgs, vector bosons, supersymmetric particles, *etc.* Here $\nu$ is the photon laboratory energy and $\Delta\sigma(\nu) = \sigma_P(\nu) - \sigma_A(\nu)$ is the difference between the photoabsorption cross section for parallel and antiparallel photon and target helicities. Similar arguments also imply that the DHG integral vanishes at tree level for virtual photoabsorption processes such as $\ell\gamma \to \ell Q\bar{Q}$ (with $Q$ being a heavy fermion) and $\ell g \to \ell Q\bar{Q}$, the lowest order sea-quark contribution to polarized deep inelastic photon and hadron structure functions. Of course the sum rule does receives individual nonzero contributions in higher order perturbation theory in the SM from quantum loop corrections and the production of higher particle number final states. However, The DHG sum rule predicts that the final result is very small, of order $\alpha$ times the square of the target's anomalous magnetic moment. The DHG sum rule thus also provides a highly non-trivial consistency check on calculations of the polarized cross sections.

In principle, one could use measurements of the logarithmic integral of the polarized photoabsorption cross section in Eq. (1) as a way to isolate the higher order radiative corrections and bound the deviations from the canonical SM couplings. Some of the most interesting applications and tests of the DHG sum rule in the SM would be to apply Eq. (1) to the reactions $\gamma\gamma \to W^+W^-$, $\gamma\gamma \to Z^0Z^0$, $\gamma e \to Ze$, and $\gamma e \to W\nu$. The delicate cancellation of the positive and negative contributions [16] of $\Delta\sigma(\gamma e \to W\nu)$ to the DHG integral calculated in Born approximation is evident in Fig. 1. On the other hand, if the $W$ were to have non-zero anomalous magnetic and electric quadrupole moments, *i.e.*, $\Delta\mu_W, \Delta Q_W \neq 0$, then the DHG integral for $\gamma e \to W\nu$ is not zero since the cancellations no longer take place [13]. In fact if the $W$ had a point-like anomalous magnetic moment, then the DHG integral for the $2 \to 2$ process diverges logarithmically at high energies.

In this paper we shall exploit the fact that the vanishing of the logarithmic integral of $\Delta\sigma$ in the Born approximation also implies that there must be a center of mass energy, $\sqrt{s}_0$, where the polarization asymmetry $A = \Delta\sigma/\sigma$ possesses a zero, *i.e.*, where $\Delta\sigma(\gamma e \to W\nu)$ reverses sign. We shall demonstrate the sensitivity of the position of this zero or 'crossing point' (which occurs at $\sqrt{s}_{\gamma e} = 3.1583\ldots M_W \simeq 254$ GeV in the SM) to modifications of the SM trilinear $\gamma WW$ coupling. As we will see, given reasonable assumptions for the luminosity and energy range for the Next Linear Collider(NLC), the zero point, $\sqrt{s_0}$, of the polarization asymmetry may be determined with sufficient precision to constrain the anomalous couplings of the $W$ to better than the 1% level at 95% CL. Since the zero occurs at rather modest energies where the unpolarized cross section is near its maximum, we will also see that an electron-positron collider with $\sqrt{s} = 320 - 400$ GeV is sufficient for our analysis, whereas other techniques [17, 18] aimed at probing the anomalous couplings through the $\gamma e \to W\nu$ process require significantly larger energies. In addition to the fact that only a limited range of energy is required, the polarization asymmetry measurements have the obvious advantage that many of the systematic errors cancel in taking cross section ratios. The position of the zero thus provides an additional weapon in the arsenal used to probe anomalous trilinear gauge couplings.

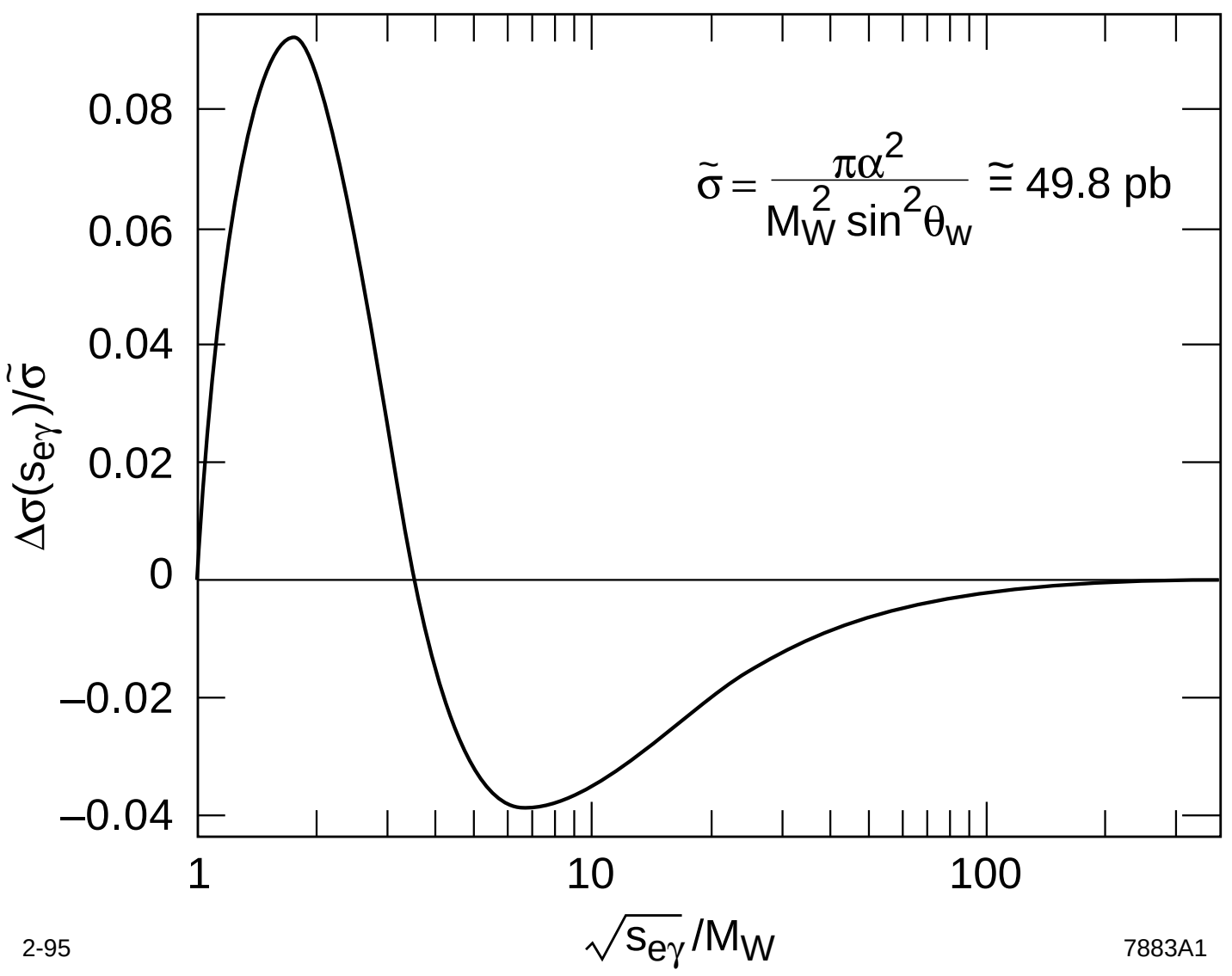

Figure 1: The Born cross section difference $\Delta\sigma$ for the Standard Model process $\gamma e \to W\nu$ for parallel minus antiparallel electron/photon helicities as a function of $y$. The logarithmic integral of $\Delta\sigma$ vanishes in the classical limit.

## II. CALCULATION OF THE POLARIZATION ASYMMETRY

The total polarization-dependent cross section for $\gamma e \to W\nu$, in the case where only the $C$- and $P$-conserving anomalous $WW\gamma$ couplings are non-zero, can be written in the form

$$\sigma = (1 - P)(\sigma_{\rm un} + \xi\sigma_{\rm pol})\,. \tag{2}$$

Here, $-1 \leq P \leq 1$ denotes the initial $e^-$ beam polarization, $\sigma_{\rm pol} = \Delta\sigma$, and the Stoke's parameter, $-1 \leq \xi \leq 1$, describes the circular polarization of the back-scattered laser photon. $\sqrt{s_{e\gamma}}$ is the center of mass energy of the $e - \gamma$ collisions. We shall consider deviations from the SM where the $W$ has point-like (momentum-independent) anomalous magnetic and quadrupole couplings. The polarization-dependent part of the cross section, $\sigma_{\rm pol}$, is then given by

$$\sigma_{\rm pol} = \frac{\tilde{\sigma}}{32x^3}\left[T_1 + (\Delta\kappa + \lambda)^2 T_2 - \Delta\kappa T_3 - \lambda T_4\right]\,, \tag{3}$$

where $x = y^2 = s_{e\gamma}/M_W^2$, $\tilde{\sigma} = \sqrt{2}G_F\alpha \simeq 49.8$ pb, and

$$\begin{aligned} T_1 &= -24 - 80x + 104x^2 - 32x(3+x)\log(x)\,, \\ T_2 &= x + 2x^2 - 3x^3 + 4x^2\log(x)\,, \\ T_3 &= 48x(1-x) + 16x(2+x)\log(x)\,, \\ T_4 &= 64x(1-x) + 32x(1+x)\log(x)\,. \end{aligned} \tag{4}$$

We have used the standard notation of Hagiwara *et al.*[19] for the anomalous static moments of the $W$: $\Delta\mu_W = \frac{e}{2M_W}(\Delta\kappa + \lambda)$ and $\Delta Q_W = \frac{-e}{M_W^2}(\Delta\kappa - \lambda)$. The corresponding polarization-independent term, $\sigma_{\rm un}$, is given by

$$\sigma_{\rm un} = \tilde{\sigma}\left[T_5 + \Delta\kappa T_6 - \lambda(\lambda + \Delta\kappa)T_7 + \lambda^2 T_8 + (\Delta\kappa + \lambda)^2 T_9\right]\,, \tag{5}$$

where

$$
\begin{aligned}
T_5 &= (1-\frac{1}{x})(1+\frac{5}{4x}+\frac{7}{4x^2})-(2+\frac{1}{x}+\frac{1}{x^2})\frac{\log(x)}{x}\,, \\
T_6 &= \frac{-1+x+2x^2}{2x^2}-\frac{2+3\ \log(x)}{2x}\,, \\
T_7 &= \frac{1}{2}\left[-1+\frac{1}{x}+\log(x)\right]\,, \\
T_8 &= \frac{(-1+x)^2}{8x}\,, \\
T_9 &= \frac{1+2x+x^2}{32x^2}-\frac{1+(1-x)\log(x)}{8x}\,.
\end{aligned}
\tag{6}
$$

These results were obtained through the use of MATHEMATICA and REDUCE and differ somewhat in their non-SM terms from other explicit results in the literature. However, the integration over angles of the helicity amplitudes obtained by Raidal [18] nicely reproduces the above expressions. Note that the effective values of $\Delta\kappa$ and $\lambda$ that are probed in the $\gamma e \rightarrow W\nu$ reaction are for *on-shell* photons and may in principle differ from those probed in $e^+e^- \rightarrow W^+W^-$ where the photon is time-like with $q^2 > 4M_W^2$.

In its $\gamma e \rightarrow W\nu$ manifestation, the DHG sum rule implies that

$$
\int_1^\infty \frac{\sigma_{\rm pol}(x)}{x}dx = 0\,, \tag{7}
$$

for the tree graph SM cross section where the couplings of all the particles involved in the process are canonical. (The electron mass is also neglected here.) In Ref. [15] it was pointed out that the vanishing of the above integral is due to a rather delicate cancellation between the regions where $\sigma_{\rm pol}$ is positive ($y = \sqrt{s_{e\gamma}}/M_W < y_0$) and regions where it is negative ($y > y_0$). Here we denote the cross-over point where the integrand vanishes (*i.e.*, the zero position) by $y_0 \simeq 3.1583$. in the SM case. When anomalous $WW\gamma$ couplings are present, several things happen. First, since the couplings are no longer canonical the DHG sum rule will be violated. Indeed, since the $\gamma e \to W\nu$ cross section is not well-behaved in the $y \to \infty$ limit when these point-like anomalous couplings are non-zero, we might also expect that the DHG integral does not even converge! This expectation is indeed realized by performing an explicit calculation employing a cut-off parameter, $x_m \gg 1$; to leading order in $x_m^{-1}$ we obtain (*i.e.*, dropping all terms of order $x_m^{-2}$ or higher)

$$\begin{aligned}\int_1^{x_m} \frac{\sigma_{\rm pol}(x)}{x} dx &= \frac{\tilde{\sigma}}{64}\Bigg[(\Delta\kappa+\lambda)^2\left(13-6\log(x_m)-8\frac{\log(x_m)}{x_m}\right) \\ &\qquad -16\lambda+64\frac{\log(x_m)}{x_m}\left(1+\lambda+\frac{1}{2}\Delta\kappa\right)\Bigg], \qquad (8)\end{aligned}$$

where we then take $x_m \to \infty$. Here we see that non-zero values of the sum $\Delta\kappa + \lambda$ result in the DHG integral becoming logarithmically divergent. (Of course, as $x$ gets large new physics effects, such as form factors and new particle production, arise to prevent the integral from truly diverging. This apparent divergence is simply the result of the break down of the point-like approximation for the anomalous couplings.) If $\Delta\mu_W = \frac{e}{2M_W}(\Delta\kappa+\lambda) = 0$ then the integral converges and yields a finite result proportional to $\lambda$. Note that the well-known radiation amplitude zero [20] (which takes place at $\cos\theta = 1$ in both the $d\sigma_{\rm pol}$ and $d\sigma_{\rm un}$ angular distributions) also occurs for this process whenever $\Delta\kappa + \lambda = 0$. Thus, if we could determine the value of the DHG integral directly from experimental data it would provide us a unique handle on possible non-zero values of $\Delta\kappa$ and $\lambda$.

In practice, the collider energy as well as the maximum energy fraction carried by the backscattered laser light are restricted. For a 500 GeV (1 TeV) $e^+e^-$ collider, only the range $1 \leq y \leq 5.4(10.4)$ is kinematically accessible. This range is far too small to allow a direct confrontation with the sum rule since we are still very far from the asymptotic region. Thus

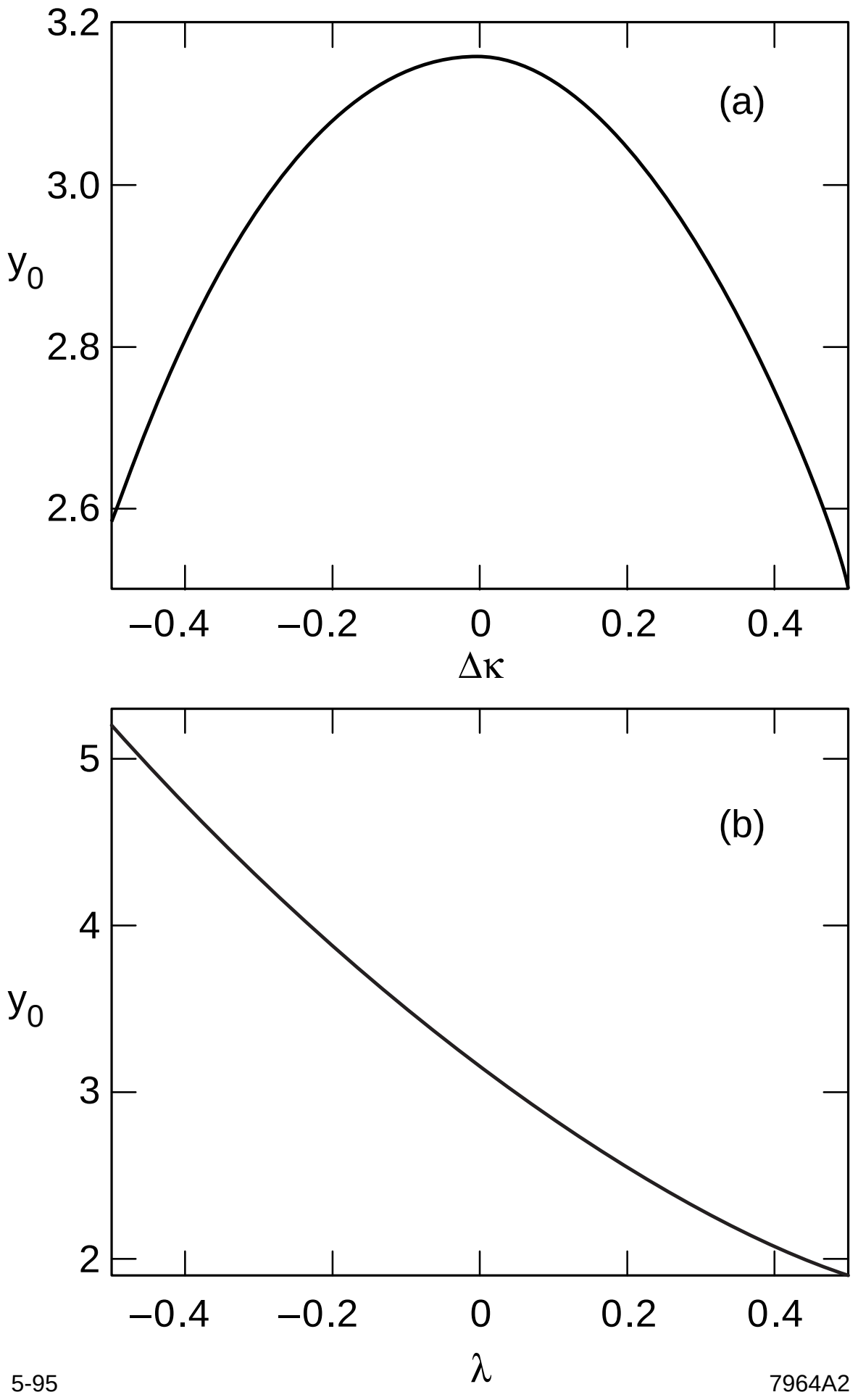

Figure 2: Separate $\Delta\kappa$ and $\lambda$ dependence of the value of $y_0$.

we must turn to more subtle methods.

Since the null result of the DHG sum rule arises from a delicate cancellation *not present* when anomalous couplings exist it is clear that these coupling must modify both the shape of $\sigma_{\rm pol}(x)$ as well as the location of the place where the integrand vanishes, $y = y_0$. Past analyses [17] have focussed on the overall shape of the polarization asymmetry, whereas here we will focus mainly on the zero's position. Figures 2a–b show the separate $\Delta\kappa$ and $\lambda$ dependence of the value of $y_0$. Several features are immediately apparent from these plots: $(i)$ If $\lambda = 0$, then the deviation of $\Delta\kappa$ from zero perturbs the value of $y_0$ to smaller values; $(ii)$ if $\Delta\kappa = 0$, the variation of $\lambda$ from zero can push $y_0$ in either direction depending on the sign of $\lambda$; $(iii)$ the value of $y_0$ shows a significantly greater sensitivity to non-zero values of $\lambda$ than $\Delta\kappa$. Thus if measurements determine that the energy where the polarization asymmetry changes sign is higher than that predicted by the SM, then $\lambda$ must be non-zero. It is also apparent that probing the location of the asymmetry zero will lead to a stronger bound on $\lambda$ than on $\Delta\kappa$.

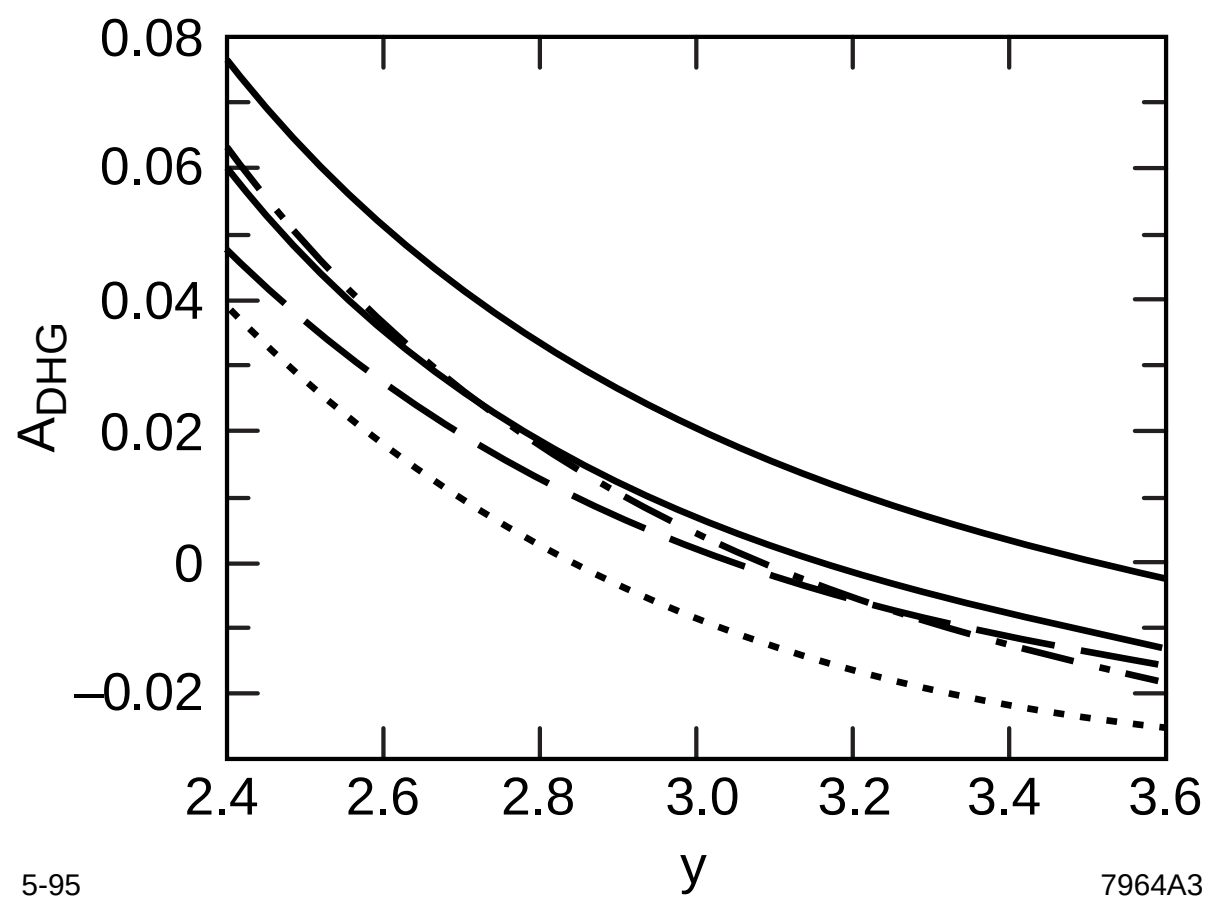

Figure 3: $A_{DHG}$ in the region near the SM value of $y_0$. The solid curve is the SM prediction while the others, from top to bottom on the left, are for $\lambda = -0.1$, $\Delta\kappa = -0.2$, $\Delta\kappa = 0.2$, $\lambda = 0.1$, respectively.

In order to ascertain how much quantitative information we can obtain on the values of $\Delta\kappa$ and $\lambda$ from measuring the crossing point $y_0$, we must perform a Monte Carlo study. Specifically, we will first want to know the constraints we can place on these anomalous coupling parameters if the SM situation is realized. We take the $y$ region surrounding the SM value of $y_0$ and divide it into 11 bins each of width $\Delta y = 0.2$. Note we have not yet tried to optimize either bin size or the distribution of integrated luminosity. Instead of considering $\sigma_{\rm pol}$, we form an asymmetry using the ratio of both cross sections

$$A_{DHG}(y) = \frac{\sigma_{\rm pol}(y)}{\sigma_{\rm un}(y)}\,, \tag{9}$$

thus removing a number of systematic errors from the analysis. Figure 3 shows that not only does the value of $y_0$ change when anomalous couplings are present, but so too does the shape of $A_{DHG}$ in the region near the zero. We assume as input into our Monte Carlo study that each $\Delta y$ bin receives an equal integrated luminosity of $5fb^{-1}$ and that the $e^-$ beam is 90% left-handed polarized, *i.e.*, $P = -0.90$. Next we generate Monte Carlo "data" (assuming the SM is correct) and try to fit the resulting distribution to the $\Delta\kappa$- and $\lambda$-dependent functional form of $A_{DHG}$. If $\lambda(\Delta\kappa)$ is zero, this procedure yields the fit shown in the first line of Table 1(with 95% CL errors). If we assume that *both* $\Delta\kappa$ and $\lambda$ non-zero, we obtain the 95% CL allowed region shown in Fig. 4. As we have anticipated, we obtain a far more restricted range of $\lambda$ than we do $\Delta\kappa$. We expect that somewhat better limits may be obtainable by optimization of our parameters. Notice that we have only performed our fit by covering the $y$ region 2.0–4.2, which could just as well have been done by an $e^+e^-$ collider with $\sqrt{s} \simeq 420$ GeV with the same integrated luminosity.

Do the constraints improve if we fit $A_{DHG}$ over the entire $y$ range accessible at a 500 GeV collider with the same total integrated luminosity? To address this question, we now take 22 bins of width $\Delta y = 0.2$ covering the range $1 \leq y \leq 5.4$ with an integrated luminosity of $2.5fb^{-1}$ per bin and re-do our fit. (We again note that we have not tried to optimize either the bin size or the distribution of integrated luminosity; we are simply seeing the sensitivity of our results to different fit assumptions.) In the case that either $\Delta\kappa$ or $\lambda$ is zero we find the fit values displayed in Table 1. The result in the case where both anomalous coupling parameters are non-zero is shown as the solid curve in Fig. 5. In either case we see no substantial improvement in the bounds we can obtain on $\lambda$ but fitting the entire accessible $y$ range *significantly* reduces the allowed range of $\Delta\kappa$ at 95% CL.

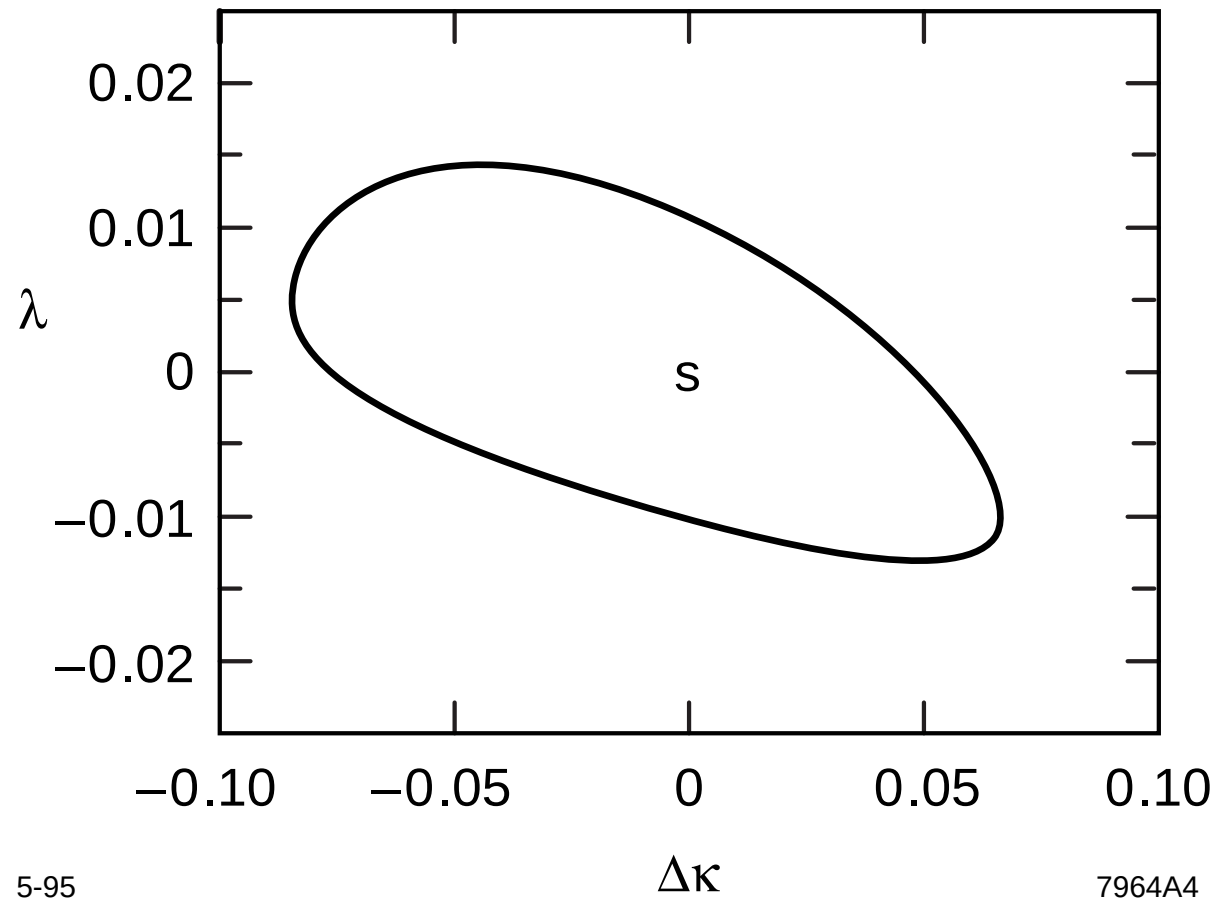

Figure 4: 95% CL in the $\Delta\kappa$-$\lambda$ plane for the 11 bin fit described in the text. 's' labels the SM prediction.

If we keep the collider energy fixed and double the integrated luminosity per bin, how do our results change for the SM example because of the improved statistics? In the case where either $\Delta\kappa$ or $\lambda$ is non-zero we find the values shown in Table 1. If both parameters are non-zero, we obtain the allowed region shown as the dashed curve in Fig. 5. The doubling of the statistics results in a significantly smaller allowed range for both of the anomalous coupling parameters.

How does the size of the bin affect these results? If we double the number of bins

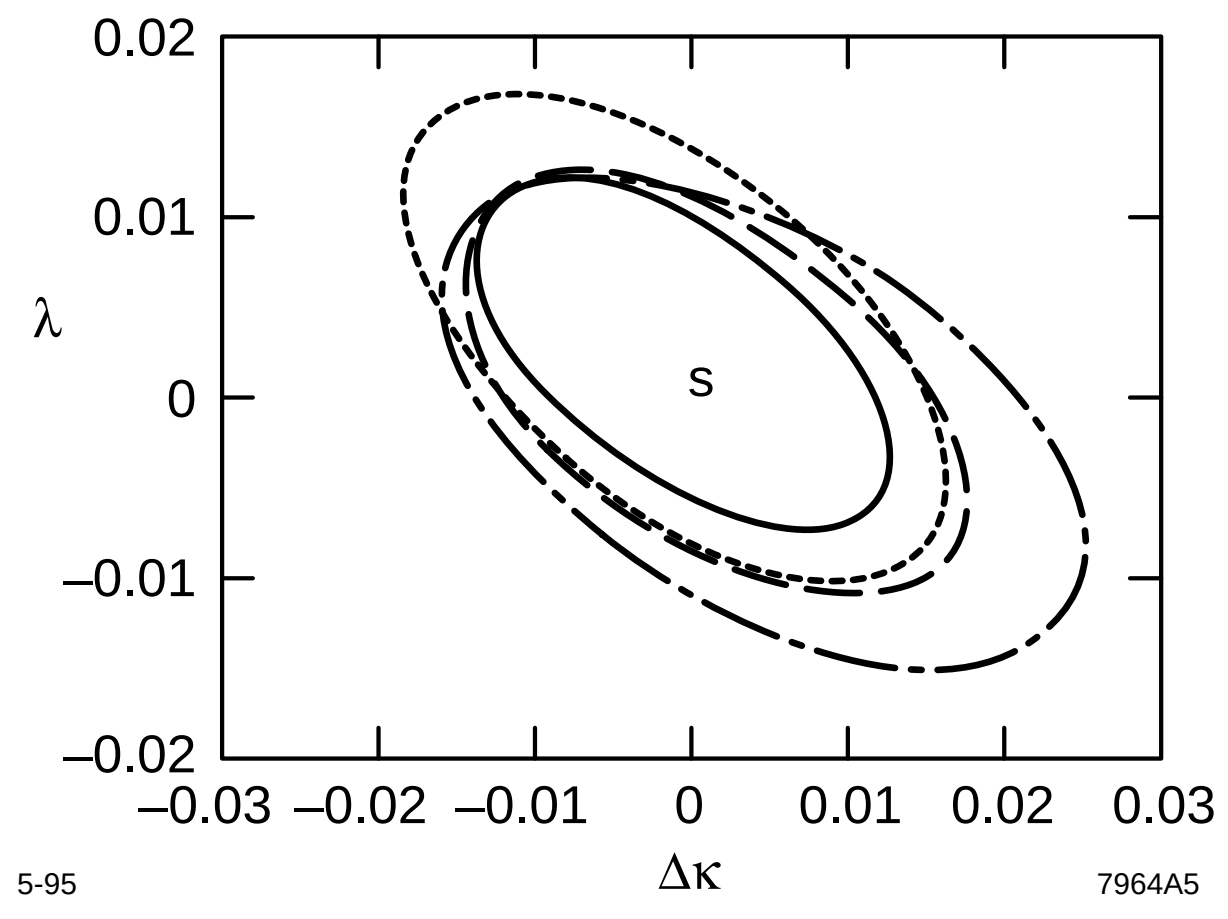

Figure 5: The solid(dashed, dash-dotted, dotted) curve corresponds to the second(third, fourth, fifth) case described in Table 1. 's' labels the SM prediction.

(*i.e.*, change to $\Delta y = 0.1$) and keep the total integrated luminosity and center of mass energy fixed for the same total $y$ range, do we improve our sensitivity? In the case where either $\Delta\kappa$ or $\lambda$ is zero we find the values shown in Table 1. The result in the case where both anomalous coupling parameters are non-zero is shown as the dash-dotted curve in Fig. 5. In this case we see a very slight degradation of the limits obtained previously but no truly significant changes.

What happens at a 1 TeV collider with higher integrated luminosity? Since fitting the entire distribution gave the best results in the 500 GeV case, we will consider only this situation. We keep the bin size and integrated luminosity per bin fixed, but extend the $y$ range up to 10.4, and repeat the above procedure. In the case that either $\Delta\kappa$ or $\lambda$ vanishes we obtain (with 95% CL errors) the values in Table 1. When both anomalous couplings are present, we obtain the 95% CL allowed region inside the dotted curve in Fig. 5. The size of the allowed region in this case is somewhat smaller than the corresponding one obtained for the 500 GeV collider but not as significantly improved as that obtained by doubling the luminosity in the 500 GeV case. A short analysis shows that essentially *all* of the improvement in the $\lambda$ determination comes from increasing the integrated luminosity whereas the improved $\Delta\kappa$ determination derives both from better statistics as well as the expanded energy range covered by the fit.

Thus given fixed integrated luminosity, the value of $\lambda$ is well constrained from fits to the data in the region near $y_0$ whereas, the optimal limits on $\Delta\kappa$ requires data fit over a large energy range.

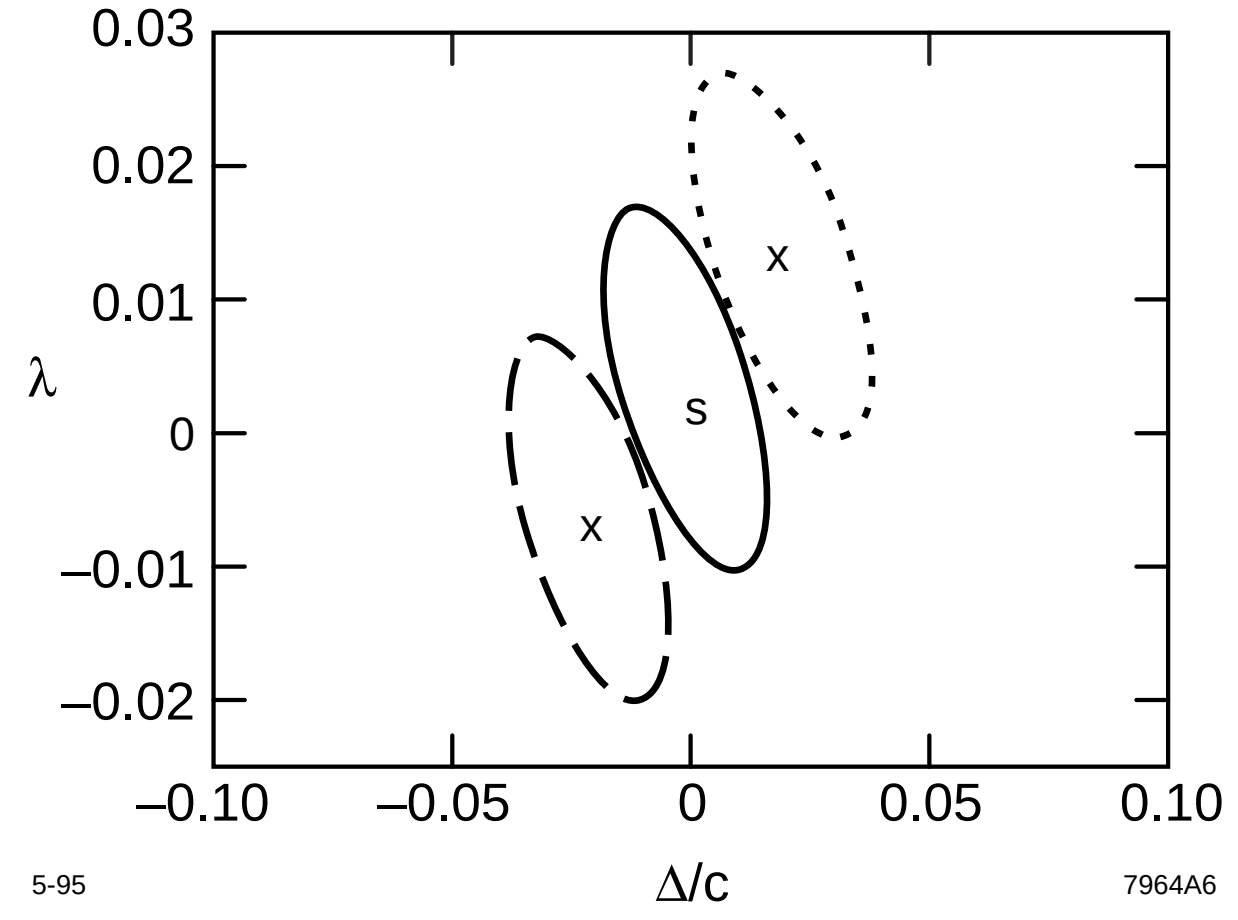

Figure 6: 95% CL regions in the $\Delta\kappa$-$\lambda$ plane for the SM case as well as for the $\Delta\kappa$ and $\lambda$ values marked by an 'x'. Note that these cases do not have overlapping allowed regions.

Let us now suppose that one or both $\Delta\kappa$ or $\lambda$, are non-zero at the percent level; can such anomalies be distinguished from the SM? For purposes of comparison, we again consider a 500 GeV machine with $\mathcal{L} = 55 fb^{-1}$ equally distributed over 22 bins with $\Delta y = 0.2$ which covers the range $1 \leq y \leq 5.4$. Figure 6 shows two simple cases with their corresponding 95% CL ellipses: (*i*) $\Delta\kappa = -0.02$ with $\lambda = -0.01$, and (*ii*) $\Delta\kappa = 0.02$ with $\lambda = 0.01$. In either case we see that the SM is excluded at 95% CL, *i.e.*, the ellipses do not overlap the SM result. The non-SM and SM cases are thus seen to be cleanly separated. The polarization asymmetry zero can be used to constrain $\Delta\kappa$ and $\lambda$ if the SM is realized, but it can also *discover* anomalous couplings if they are indeed present.

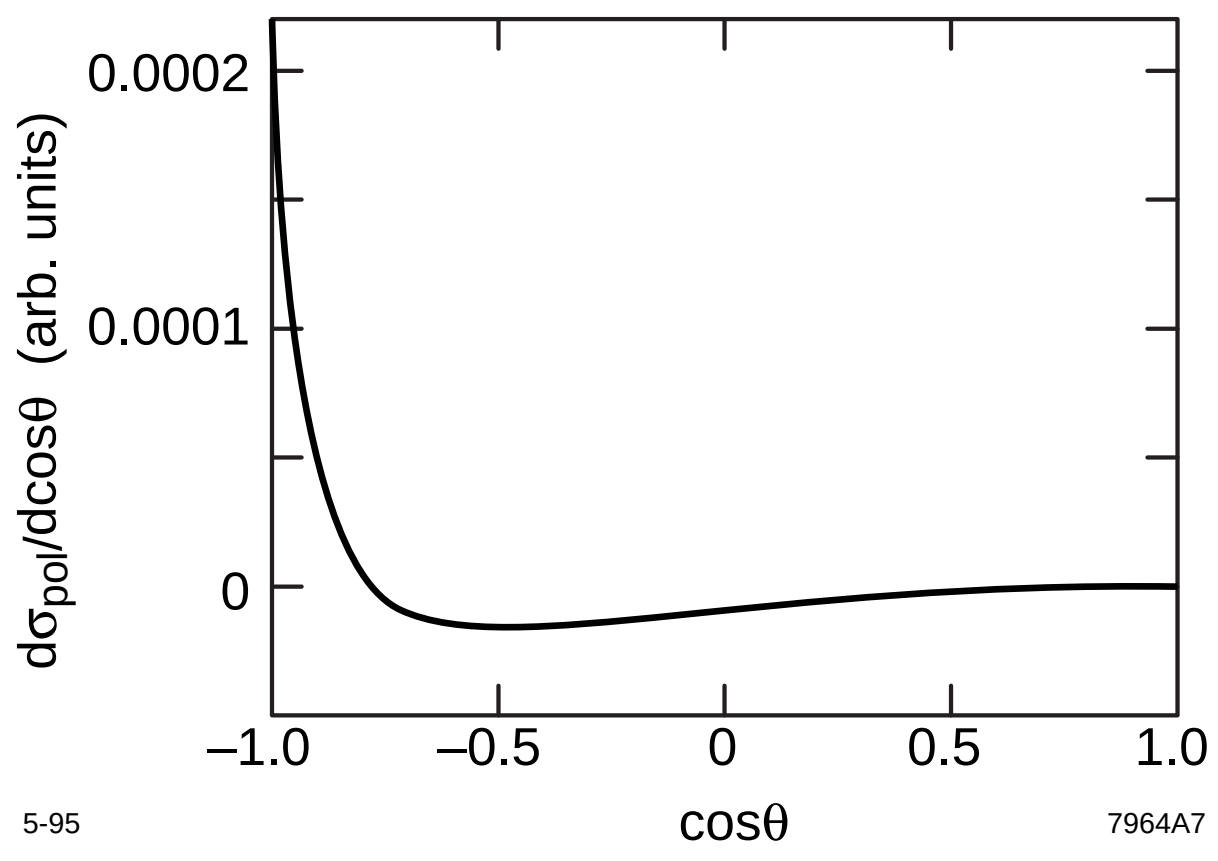

Figure 7: Angular distribution of the polarization-dependent part of the $\gamma e \rightarrow W\nu$ cross section in the SM for $y = y_0$. Note the radiation zero at $\cos\theta = 1$ and that the integral over $\cos\theta$ vanishes.

Thus far we have only considered an equal distribution of integrated luminosity in a fairly wide region surrounding $y_0$. What happens in the reverse scenario, *i.e.*, when we concentrate all of our luminosity in a single bin surrounding the SM value of $y_0 \simeq 3.158$? In this case an $e^+e^-$ collider with an energy of only $\sqrt{s} \simeq 320$ GeV is needed. From the previous discussion we expect to obtain a relatively very poor constraint on $\Delta\kappa$ since the energy range is so limited. As a benchmark, let us again take $\mathcal{L} = 55fb^{-1}$ and consider bin widths of $\Delta y = 0.1,\ 0.2,$ and 0.3. If $\Delta\kappa = 0$, we find in all three cases that $\lambda = -0.0009 \pm 0.0086$, whereas if $\lambda = 0$ we obtain instead $\Delta\kappa = -0.028^{+0.107}_{-0.095}$. As anticipated, $\Delta\kappa$ is relatively poorly determined, but $\lambda$ is as well-constrained as in the case of the wide range 11 or 22 bin fits. Thus an excellent bound on $\lambda$ can be achieved even with an $e^+e^-$ collider of rather modest center of mass energy ($\simeq$ 320 GeV); achieving strong limits on $\Delta\kappa$ will requires a higher energy machine.

As a last case, we take the same integrated luminosity as above and reintroduce 11,

$\Delta y = 0.2$ bins. However, instead of giving all 11 the same luminosity, we preferentially weight the bins closest to the central bin where the SM value predicts the change in sign of the polarized asymmetry. As an example, we take 1/2 of the total luminosity in the $y = 3.05$–3.25 central bin, 1/8 in the two adjacent bins, 1/16 in the next pair and so on until the $y$ range 2.05–4.25 is covered. (The central bin thus receives a luminosity of $\mathcal{L} = 55 \cdot \frac{1}{2} \cdot \frac{64}{63}\ fb^{-1}$.) In this case we find that $\Delta\kappa = 0.0165^{+0.0657}_{-0.1313}$ when $\lambda = 0$ (a *very* poor determination, as expected) and $\lambda = -0.0018^{+0.0080}_{-0.0085}$, when $\Delta\kappa = 0$. Note that this $\lambda$ range is quite comparable to that found in all the other cases above with the same total luminosity. If we changed the initial weighting fraction of 1/2 for the central bin to $1/f$, with $f > 1$, then we obtain the results displayed in Table 2. In all cases the $\Delta\kappa$ determination is quite poor while $\lambda$ is well determined.

At the crossing point of the polarization asymmetry, the integral over angles of the of the polarization-dependent part of the $\gamma e \to W\nu$ cross section vanishes. By the mean value theorem there must be an angle $\theta^0_{\rm cm}$ where the polarization-dependent part of the differential cross section vanishes. This is illustrated in Fig. 7 for the case of the SM. In principle, measurements of this angular zero can also be used to limit possible new physics and anomalous couplings of the $W$, but the limited angular acceptance will cause a corresponding detriment to the statistical significance.

## III. DISCUSSION AND CONCLUSIONS

There has been a remarkable progression of improvement in the absolute precision of the measurements for the lepton magnetic moments – to parts in $10^{-8}$ for the electron and parts per $10^{-6}$ for the muon [21]. The lepton $g-2$ measurements exploit the fact that the canonical Dirac coupling implies the equality of the Larmor and spin-precession frequencies of charged particles of any non-zero spin in a constant magnetic field. Unfortunately, it is unlikely that this precise technique could be directly exploited to determine the moments of particles as short lived as the vector bosons of the standard model.

In this paper we have exploited two other unique features implied by canonical magnetic and electric quadrupole couplings: (1) the fact that the logarithmic integral of the polarized photoabsorption cross section vanishes identically, and the corollary (2) that there exists a specific energy where the polarization asymmetry must reverse sign.

In particular, we have shown that measurements of the DHG integral for the process $\gamma e^- \to W^- \nu$ can bound the anomalous magnetic moment of the $W$. The polarization asymmetry for this process is particularly sensitive to the anomalous coupling $\Delta\kappa$ at high energies. The SM also predicts that polarization asymmetry for $\gamma e^- \to W^- \nu$ vanishes at a precise energy $\sqrt{s_{\gamma e}} = 3.1583\ldots M_W \simeq 254$ GeV in Born approximation. We have shown that measurements of any deviation from this value can provide 95% confidence level limits on the $W$ coupling $\lambda$ at a precision significantly below 1%, assuming the projected NLC luminosity. The precision of this polarization asymmetry measurement will benefit from the relatively low linear collider energy and the smaller systematic errors associated with asymmetries.

## IV. ACKNOWLEDGMENTS

We would like to thank J.L. Hewett for discussions related to this work. One of us (TGR) also thanks the members of both the Phenomenology Institute at the University of Wisconsin-Madison and the Argonne National Laboratory High Energy Theory Group for the use of their computing facilities as well as their hospitality.

| $\sqrt{s}(GeV)$ | $\mathcal{L}$/bin $fb^{-1}$ | $N_{\rm bins}$ | $\Delta\kappa$ | $\lambda$ |
|---|---|---|---|---|
| 500 | 5 | 11 | $-0.008^{+0.046}_{-0.054}$ | $0.0004^{+0.0085}_{-0.0086}$ |
| 500 | 2.5 | 22 | $0.002^{+0.011}_{-0.012}$ | $0.0029 \pm 0.0086$ |
| 500 | 5 | 22 | $0.0014^{+0.0087}_{-0.0086}$ | $0.0022^{+0.0062}_{-0.0063}$ |
| 500 | 1.25 | 44 | $0.0037^{+0.0139}_{-0.0140}$ | $0.0005 \pm 0.0092$ |
| 1000 | 2.5 | 47 | $0.002 \pm 0.011$ | $0.0012^{+0.0078}_{-0.0080}$ |

Table 1: 95% CL constraints on $\Delta\kappa$ and $\lambda$ for the different scenarios described in the text.

| $f$ | $\lambda$ |
|---|---|
| 1.1 | $-0.0008^{+0.0087}_{-0.0085}$ |
| 1.5 | $-0.0018^{+0.0087}_{-0.0085}$ |
| 2.0 | $-0.0018^{+0.0080}_{-0.0085}$ |
| 3.0 | $-0.0015^{+0.0087}_{-0.0085}$ |
| 4.0 | $-0.0012^{+0.0086}_{-0.0086}$ |

Table 2: 95% CL constraints on $\lambda$ for the different values of the parameter $f$ described in the text.